\documentclass[sts,preprint]{imsart} 
\usepackage{graphicx}
\usepackage{amssymb, amsmath, amsthm} 
\usepackage{epstopdf} 
\usepackage{subfigure} 
\usepackage{natbib} 
\usepackage[colorlinks,citecolor=blue,urlcolor=blue]{hyperref} 
\usepackage{listings}
\usepackage{algorithm}
\usepackage[noend]{algpseudocode}

\graphicspath{{figures/}} 

\usepackage{color}

\usepackage[usenames,dvipsnames,svgnames,table]{xcolor}
\definecolor{codegreen}{rgb}{0,0.6,0}
\definecolor{codegray}{rgb}{0.5,0.5,0.5}
\definecolor{codepurple}{rgb}{0.58,0,0.82}
\definecolor{backcolour}{rgb}{0.99,0.99,0.97}
\lstdefinestyle{stan}{
	literate={~}{$\sim$}{1},
	backgroundcolor=\color{backcolour},
	commentstyle=\color{codegreen},
	morekeywords = {real, vector, matrix, data, model, parameters, transformed, lower, generated, quantities},
	keywordstyle=\color{magenta},
	numberstyle=\tiny\color{codegray},
	stringstyle=\color{codepurple},
	emph={%
		normal, cauchy, inv_gamma, bernoulli_logit, gamma, normal_rng%
	},
	emphstyle=\color{codepurple},%
	basicstyle={\footnotesize \ttfamily},
	breakatwhitespace=false,
	breaklines=true,
	captionpos=b,
	keepspaces=true,
	numbers=left,
	numbersep=5pt,
	showspaces=false,
	showstringspaces=false,
	showtabs=false,
	tabsize=2
}
\newcommand{\norm}[1]{\left\|#1\right\|}

\newtheorem{theorem}{Theorem}

\begin{document}

\begin{frontmatter}

\title{Validating Bayesian Inference Algorithms with Simulation-Based Calibration}
\runtitle{Simulation-Based Calibration}

\begin{aug}
  \author{Sean Talts%
  \ead[label=e1]{sean.talts@gmail.com}},
  \author{Michael Betancourt%
  \ead[label=e2]{betan@symplectomorphic.com}},
  \author{Daniel Simpson%
  \ead[label=e3]{simpson@utstat.toronto.edu}},
  \author{Aki Vehtari%
  \ead[label=e4]{Aki.Vehtari@aalto.fi}},
  \author{Andrew Gelman%
  \ead[label=e5]{gelman@stat.columbia.edu}}

  \runauthor{S. Talts \emph{et al.}}

  \address{ISERP, Columbia University, New York. \printead{e1}.}
  \address{Symplectomorphic LLC., New York. \printead{e2}.}
  \address{Department of Statistical Sciences, University of
           Toronto, Toronto. \printead{e3}.}
  \address{Department of Computer Science, Aalto University,
           Finland. \printead{e4}.}
  \address{Department of Statistics and Department of
           Political Science, Columbia University, New York.
           \printead{e5}.}

\end{aug}

\begin{abstract}
Verifying the correctness of Bayesian computation is challenging. This is
especially true for  complex models that are common in practice, as these require
sophisticated model implementations and algorithms.  In this paper
we introduce \emph{simulation-based calibration} (SBC), a general
procedure for validating inferences from Bayesian algorithms
capable of generating posterior samples. This procedure not only
identifies inaccurate computation and inconsistencies in model
implementations but also provides graphical summaries that can
indicate the nature of the problems that arise.  We argue that SBC
is a critical part of a robust Bayesian workflow, as well as being
a useful tool for those developing computational algorithms and
statistical software.
\end{abstract}

\end{frontmatter}

\section{Introduction}

Powerful algorithms and computational resources are facilitating
Bayesian modeling in an increasing range of applications.
Conceptually, constructing a Bayesian analysis is straightforward.
We first define a joint distribution over the parameters, $\theta$,
and measurements, $y$, with the specification of a prior distribution
and likelihood,
\begin{equation*}
\pi(y, \theta) = \pi(y \mid \theta) \, \pi(\theta).
\end{equation*}
Conditioning this joint distribution on an observation, $\tilde{y}$,
yields a posterior distribution,
\begin{equation*}
\pi(\theta \mid \tilde{y}) \propto \pi(\tilde{y}, \theta),
\end{equation*}
that encodes information about the system being analyzed.

Implementing this Bayesian inference in practice, however, can be
computationally challenging when applied to large and structured
datasets. We must make our model rich enough to capture the relevant
structure of the system being studied while simultaneously being
able to accurately work with the resulting posterior distribution.
Unfortunately, every algorithm in computational statistics requires
that the posterior distribution possesses certain favorable properties
in order to be successful. Consequently the overall performance of
an algorithm is sensitive to the details of the model and the
observed data, and an algorithm that works well in one analysis
can fail spectacularly in another.

As we move towards creating sophisticated, bespoke models with
each analysis, we stress the algorithms in our statistical
toolbox. Moreover, the complexity of these models provides
abundant opportunity for mistakes in their specification.
We must verify both that our code is implementing the model we
think it is and that our inference algorithm is able to perform
the necessary computations accurately. While we always get
some result from a given algorithm, we have no idea how good
it might be without some form of validation.

Fortunately, the structure of the Bayesian joint distribution
allows for the validation of \emph{any} Bayesian computational
method capable of producing samples from the posterior
distribution, or an approximation thereof. This includes
not only Monte Carlo methods but also deterministic
methods that yield approximate posterior distributions
amenable to exact sampling, such as integrated nested Laplace
approximation (INLA) \citep{rue2009approximate,rue2017bayesian}
and automatic differentiation variational inference (ADVI)
\citep{kucukelbir2017automatic}. In this paper we introduce
\emph{Simulation-Based Calibration} (SBC), a corrected implementation of the ideas of \cite{Cook2006-se} for validating these algorithms in a generic and straightforward way
within the scope of a given Bayesian joint distribution.

We begin with a discussion the natural self-consistency of
samples from the Bayesian joint distribution and previous
validation methods that have exploited this behavior. Next
we introduce the simulation-based calibration framework and
examine the qualitative interpretation of the SBC output, how
it identifies how the algorithm being validated might be failing,
and how it can be incorporated into a robust Bayesian workflow.
Finally, we consider some useful extensions of SBC before
demonstrating the application of the procedure over a range
of analyses.

\section{Self-Consistency of the Bayesian Joint Distribution}

The most straightforward way to validate a computed posterior
distribution is to compare computed expectations with the exact
values. An immediate problem with this, however, is that we
know the true posterior expectation values for only the simplest
models. These simple models, moreover, typically have a
different structure to the models of interest in applications.
This motivates us to construct a validation procedure that does
not require access to the exact expectations, or any other
property of the true posterior distribution.

A popular alternative to comparing the computed and true
expectation values directly is to define a ground truth
$\tilde{\theta}$, simulate data from that ground truth,
$\tilde{y} \sim \pi(y \mid \tilde{\theta})$, and then quantify
how well the computed posterior recovers the ground truth in
some way. Unfortunately this approach is flawed, as demonstarted
in a simple example.

Consider the model
\begin{align*}
y &\mid \mu \sim \mbox{N} (\mu, 1^{2}) \\
\mu &\sim \mbox{N} (0, 1^{2}),
\end{align*}
and an attempt at verification that uses the single ground
truth value $\tilde{\mu} = 0$. If we simulate from this model
and draw the plausible, but extreme, data value $\tilde{y} = 2.1$,
then the true posterior will be
$\mu \mid \tilde{y} \sim \mbox{N} (1.05, 0.5^2)$. As $\tilde{\mu}$
is more than two posterior standard deviations from the posterior
mean, we might be tempted to say that recovery has not been
successful. On the other hand, imagine that we accidentally used
code that exactly fits an identical model but with the variance
for both the likelihood and prior set to 10 instead of 1. In this
case, the incorrectly computed posterior would be $\mbox{N}(1.05,5^2)$
and we might conclude that the code correctly recovered the posterior.

Consequently, the behavior of the algorithm in any \emph{individual}
simulation will not characterize the ability of the inference
algorithm to fit that particular model in any meaningful way. In
the example above, it might lead us to conclude that the incorrectly
coded analysis worked as desired, while the correctly coded
analysis failed.  In order to properly characterize an analysis
we need to at the very least consider multiple ground truths.

Which ground truths, however, should we consider?  An algorithm
might be able to recover a posterior constructed from data generated
from some parts of the parameter space while faring poorly on data
generated from other parts of parameter space. In Bayesian inference
a proper prior distribution quantifies exactly which parameter values
are relevant and hence should be considered when evaluating an
analysis. This immediately suggests that we consider the performance
of an algorithm over the entire Bayesian joint distribution, first
sampling a ground truth from the prior,
$\tilde{\theta} \sim \pi(\theta)$, and then data from the
corresponding data generating process,
$\tilde{y} \sim \pi(y \mid \tilde{\theta})$. We can then
build inferences for each simulated observation $\tilde{y}$
and then compare the recovered posterior distribution to the
sampled parameter $\tilde{\theta}$.

Advantageously, this procedure also defines a natural condition
for quantifying the faithfulness of the computed posterior
distributions, regardless of the structure of the model itself.
Integrating the exact posteriors over the Bayesian joint
distribution returns the prior distribution,
\begin{equation} \label{eqn:marginal_post}
\pi(\theta) = \int
\mathrm{d} \tilde{y} \, \mathrm{d} \tilde{\theta} \,
\pi(\theta \mid \tilde{y}) \,
\pi( \tilde{y} \mid \tilde{\theta}) \,
\pi( \tilde{\theta}).
\end{equation}
In other words, for \emph{any} model the average of any
exact posterior expectation with respect to data generated
from the Bayesian joint distribution reduces to the
corresponding prior expectation.

Consequently, any discrepancy between the \emph{data averaged
posterior} \eqref{eqn:marginal_post} and the prior
distribution indicates some error in the Bayesian analysis.
This error can come either from inaccurate computation of
the posterior or a mis-implementation of the model itself.
Well-defined comparisons of these two distributions then
provides a generic means of validating the analysis, at
least within the scope of the modeling assumptions.

\section{Existing Validation Methods Exploiting the Bayesian Joint Distribution}
\label{sec:prior_work}

The self-consistency of the data-averaged posterior
\eqref{eqn:marginal_post} and the prior is not a novel
observation. This behavior has been exploited in at least
two earlier methods for validating Bayesian computational
algorithms.

\citet{Geweke2004-hr} proposed a Gibbs sampler targeting
the Bayesian joint distribution that alternatively samples
from the posterior, $\pi(\theta \mid y)$, and the likelihood,
$\pi(y \mid \theta)$. If an algorithm can generate accurate
posterior samples, then this Gibbs sampler will produce
accurate samples from the Bayesian joint distribution, and
the marginal parameter samples will be indistinguishable from
any sample of the prior distribution. The author recommended
quantifying the consistency of the marginal parameter samples
and a prior sample with $z$-scores of each parameter mean,
with large $z$-scores indicating a failure of the algorithm
to produce accurate posterior samples.

The main challenge with this method is that the diagnostic
$z$-scores will be meaningful only once the Gibbs sampler
has converged. Unfortunately, the data and the parameters will
be strongly correlated in a generative model and the convergence
of this Gibbs sampler will be slow, making it challenging to
identify when the diagnostics can be considered.

\cite{Cook2006-se} avoided the auxiliary Gibbs sampler entirely
by considering cumulative distribution function (CDF)
values (quantiles) approximated using
samples from the simulated posterior distribution.  They use the notation $\theta$ to represent any scalar model parameter or function of parameters.
They noted that if $\tilde{\theta} \sim \pi(\theta)$ and
$\tilde{y} \sim \pi(y \mid \tilde{\theta})$ then the exact
posterior CDF values for each parameter,
\begin{equation*}
q(\tilde{\theta}) =
\int \mathrm{d} \theta \, \pi ( \theta \mid \tilde{y} )
\, \mathbb{I} [ \theta < \tilde{\theta} ],
\end{equation*}
will be uniformly distributed provided that the posteriors are
absolutely continuous. Consequently any deviation from the
uniformity of the computed posterior CDF values indicates a
failure in the implementation of the analysis.

The authors suggest quantifying the uniformity of these CDF values
by transforming them into $z$-scores with an application
of the inverse normal CDF. The
absolute value of the $z$-scores can then be visualized to
identify deviations from normality of, and hence uniformity
of the CDF values. At the same time these deviations can be
quantified with a $\chi^{2}$ test.

This procedure works well in certain examples, as demonstrated
by \cite{Cook2006-se}, but it can run into problems with
MCMC samples as the empirical CDF values
only asymptotically approach the true values.
Without a central limit theorem and sufficiently small autocorrelations,
the estimated quantiles from finite MCMC samples will not follow the
uniform distribution assumed in \cite{Cook2006-se}.
These issues make it difficult to determine whether a deviation
from normality is due to pre-asymptotic behavior or biases
in the posterior computations. In addition the description
of the algorithm in \cite{Cook2006-se} is incomplete in that
it neglected to mention the continuity correction used for
its quantile computation, as implemented in \cite{Cook2006validate}.

In particular, because there are only $L+1$ positions in a
posterior sample of size $L$ in between which the prior sample
$\tilde{\theta}$ can fall, an empirically approximated CDF value
of the prior draw $\tilde{\theta}$ within the posterior sample $\theta$,
\begin{equation*}
  q = \frac{1}{L} \sum_{l=1}^L \mathbb{I}[\theta < \tilde{\theta}],
\end{equation*}
is fundamentally
discrete, taking one of $L+1$ evenly spaced values on $[0, 1]$.
This discretization causes artifacts when visualizing the
CDF values and it requires some continuity corrections for the finite
instances where the estimated CDF value equals 0 or 1.
At the same time, autocorrelation in the simulations creates
dependence in the estimated CDF values and modifies the distributions of
test statistics that were worked out implicitly assuming
independence, a point recognized in the recent correction
\citep{Cook2006-se-correction}. With attempts at smoothing,
we may fix visual artifacts but we have found no exact proofs
of distribution for these continuous estimators.

\begin{figure}
  \centering
  \begin{minipage}{0.45\textwidth}
    \centering
    \includegraphics[width=\textwidth]{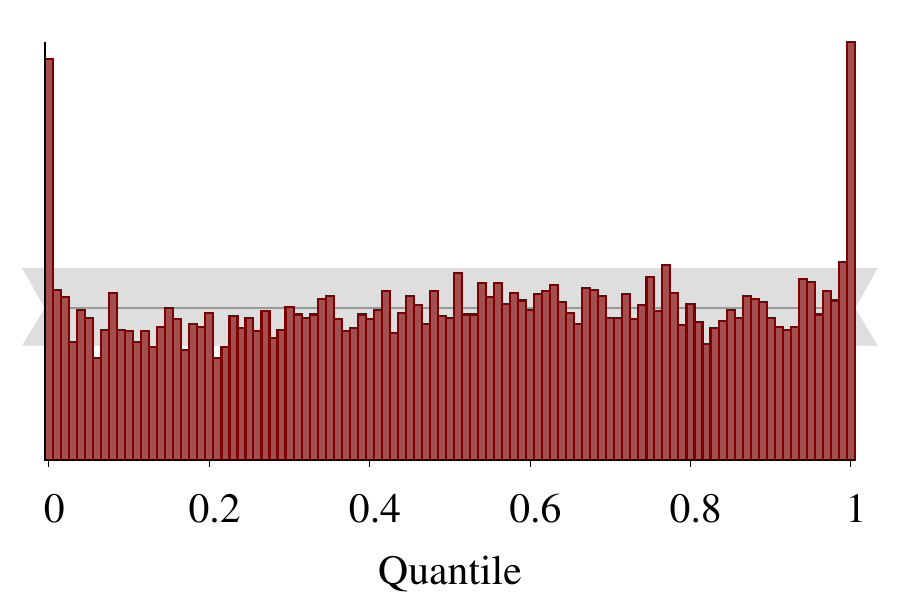}
    \caption{The procedure of \cite{Cook2006-se} applied to a linear regression
             analysis with Stan indicates significant problems despite the
             analysis itself being correct. In particular, the histogram of
             estimated CDF values (red) exhibits strong systematic deviations
             from the variation expected of a uniform histogram (gray).}
    \label{fig:lin_regr_c1}
  \end{minipage}\hfill
  \begin{minipage}{0.45\textwidth}
    \centering
    \includegraphics[width=\textwidth]{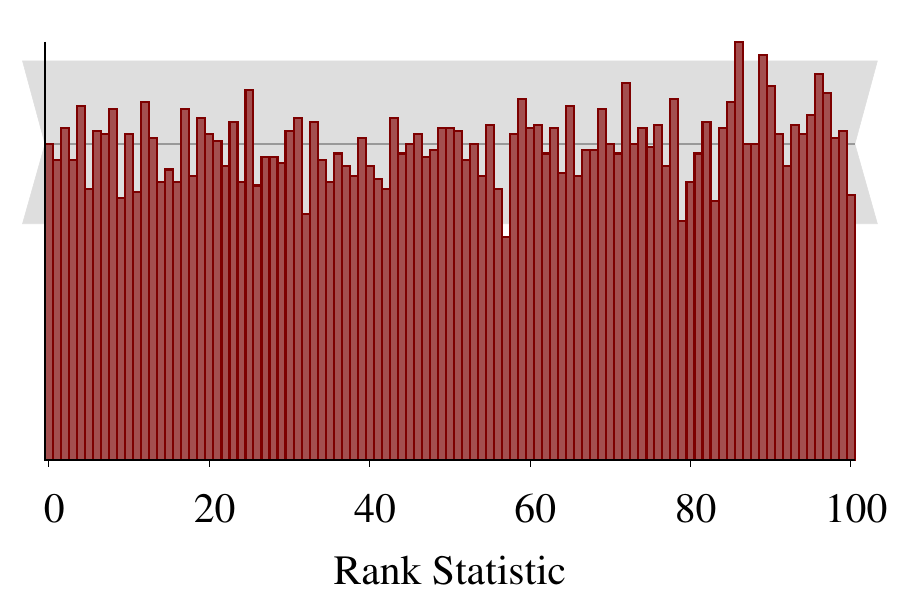}
    \caption{SBC Algorithm \ref{algo:sbc_mcmc} applied to a linear regression
             analysis indicates no issues as the empirical rank statistics (red)
             are consistent with the variation expected of a uniform histogram
             (gray).}
    \label{fig:sbac_lin_regr}
  \end{minipage}\hfill
\end{figure}

To demonstrate these issues, we run most of the
\cite{Cook2006-se} procedure for a straightforward linear
regression model
(Listings \ref{lst:gen_lin_regr_c} and \ref{lst:linregr_c} in the Appendix) in Stan 2.17.1
\citep{JSSv076i01}. The
$\Phi^{-1}$ transformation is not defined at 0 or 1, a problem with the
underlying framework that was approximately avoided in \cite{Cook2006validate}
by adding an offset 0.5 to the estimated quantiles as a continuity correction \citep{Blom:1958}.
Here, we avoid the need for continuity corrections entirely
by visualizing the estimated quantiles with a carefully-binned histogram.
For both plots in Figure
\ref{fig:lin_regr_c1}, we generated 10,000
draws from the prior predictive $\pi(y)$ and fit the Stan model on each of
these, taking 100 post-warmup draws from the posterior for each draw from the
prior predictive. For this evaluation, we used a histogram of both $\alpha$ and
$\beta$ parameters together in the same plot, as it was already evident that
non-uniformities had been found from the combined plot.

Although Stan is known to be extremely
accurate for this analysis, a histogram of the empirical
CDF values demonstrates strong deviations from uniformity
(Figure \ref{fig:lin_regr_c1}) that immediately suggests
algorithmic problems that aren't there. We also see evidence of autocorrelation
in the posterior sample manifesting in the histogram, an issue
we consider more thoroughly in Section \ref{dealAutoCor}.


\section{Simulation-Based Calibration}
\label{sec:SBC}

We can work around the discretization artifacts of \cite{Cook2006-se}
by considering a similar consistency criterion that is
immediately compatible with sampling-based algorithms. In this
section we introduce \emph{simulation-based calibration} (SBC) based on
comparing histograms of \emph{rank statistics} to
the discrete uniform distribution that would arise if the analysis has
been correctly implemented.

SBC requires just one assumption: that we have a generative model
for our data. Given such a model, we can run any given algorithm
over many simulated observations and the self consistency condition
\eqref{eqn:marginal_post} provides a target to verify that the
algorithm is accurate over that ensemble, and hence sufficiently
\emph{calibrated} for the assumed model.
This calibration ensures that certain one
dimensional test statistics are correctly distributed under the assumed model
and is similar to checking the coverage of a
credible interval under the assumed model.

Importantly, this calibration is limited exclusively to the
computational aspect of our analysis. It offers no guarantee
that the posterior will cover the ground truth for any single
observation or that the model will be rich enough to capture
the truth at all. Understanding the range of posterior behaviors
for a given observation requires a more careful
\emph{sensitivity analysis} while validating the model assumptions
themselves requires a study of \emph{predictive performance},
such as posterior predictive checks (PPCs, e.g., \cite{bda3}, chapter 6).
Where SBC uses draws from the
joint prior distribution $\pi(\theta, y)$, PPCs use the posterior
predictive distribution for predicting new data $\tilde{y}$, $\pi(\tilde{y} | y)$.
We view both of these checks as a vital part of a robust Bayesian workflow.

In this section we first demonstrate the expected behavior of
rank statistics under a proper analysis and construct the SBC
procedure to exploit this behavior. We then demonstrate how
deviations from the expected behavior are interpretable and
help identify the exact nature of implementation error.

\subsection{Validating Consistency With Rank Statistics}

Consider the sequence of samples from the Bayesian joint
distribution and resulting posteriors,
\begin{align} \label{eqn:joint_sample}
\tilde{\theta} &\sim \pi(\theta) \nonumber
\\
\tilde{y} &\sim \pi(y \mid \tilde{\theta}) \nonumber
\\
\left\{ \theta_{1}, \ldots, \theta_{L} \right\} &\sim \pi(\theta \mid \tilde{y}).
\end{align}
The relationship \eqref{eqn:marginal_post} implies that the
prior sample, $\tilde{\theta}$, and an exact posterior sample,
$\left\{ \theta_{1}, \ldots, \theta_{L} \right\}$,
will be distributed according to the the same distribution.
Consequently, for any one-dimensional random variable,
$f : \Theta \rightarrow \mathbb{R}$, the \emph{rank statistic}
of the prior sample relative to the posterior sample,
\begin{equation*} \label{eqn:rank_stat}
r(\{f(\theta_1), \ldots, f(\theta_L) \}, f(\tilde{\theta}))
= \sum_{l = 1}^{L} \mathbb{I} [ f(\theta_{l}) < f(\tilde{\theta}) ]
\in [0, L],
\end{equation*}
will be uniformly distributed across the integers $[0, L]$.

\begin{theorem} \label{theorem1}
Let $\tilde{\theta} \sim \pi(\theta)$, $\tilde{y} \sim \pi(y \mid \tilde{\theta})$,
and $\left\{ \theta_{1}, \ldots, \theta_{L} \right\} \sim \pi(\theta \mid \tilde{y})$
for any joint distribution $\pi(y, \theta)$. The rank statistic of any one-dimensional
random variable over $\theta$ is uniformly distributed over the integers $[0, L]$.
\end{theorem}

\noindent The proof is given in Appendix \ref{sec:proof}.

There are many ways of testing the uniformity of the rank statistics,
but the SBC procedure, outlined in Algorithm \ref{algo:sbc_nominal},
exploits a histogram of rank statistics for a given random variable
to enable visual inspection of uniformity (Figure \ref{fig:rank_good}).
We first sample $N$ draws from the Bayesian joint distribution.
For each replicated generated dataset we then sample $L$ exact draws from
the posterior distribution and compute the corresponding rank statistic.
We then bin the $L$ rank statistics in a histogram spanning the $L + 1$
possible values, $\left\{0, \ldots, L\right\}$.  If only correlated
posteriors samples can be drawn then the procedure can be modified
as discussed in Section \ref{dealAutoCor}.

In order to help identify deviations, each histogram is complemented
with a gray band indicating 99\% of the variation expected from a
uniform histogram. Formally, the vertical extent of the band extends
from the 0.005 percentile to the 0.995 percentile of the
$\mathrm{Binomial}(N, (L + 1)^{-1})$ distribution so that under uniformity
we expect that, on average, the counts in only one bin in a hundred will
deviate outside this band.

\begin{algorithm}[t]
\caption{SBC generates a histogram from an ensemble of rank statistics
of prior samples relative to corresponding posterior samples. Any
deviation from uniformity of this histogram indicates that the posterior
samples are inconsistent with the prior samples. For a multidimensional
problem the procedure is repeated for each parameter or quantity of
interest to give multiple histograms.}
\label{algo:sbc_nominal}
\begin{algorithmic}[0]
\State Initialize a histogram with bins centered around $0, \ldots, L$.
\State
\For{$n$ in $N$}
  \State Draw a prior sample, $\tilde{\theta} \sim \pi(\theta)$
  \State Draw a simulated data set, $\tilde{y} \sim \pi(y \mid \tilde{\theta})$
  \State Draw posterior samples
  $\left\{ \theta_{1}, \ldots, \theta_{L} \right\} \sim \pi(\theta \mid \tilde{y})$
  \For{each one-dimensional random variable, $f$}
      \State Compute the rank statistic
             $r(\{f(\theta_1), \ldots, f(\theta_L)\}, f(\tilde{\theta}))$
             as defined in \eqref{eqn:rank_stat}
      \State Increment the histogram with
             $r(\{f(\theta_1), \ldots, f(\theta_L)\}, \mid f(\tilde{\theta}))$
  \EndFor
\EndFor
\State Analyze the histogram for uniformity.
\end{algorithmic}
\end{algorithm}

In complex problems computational resources often limit the number
of replications, $N$, and hence the sensitivity of the resulting
SBC histogram. In order to reduce the noise from small replications
it can be beneficial to uniformly bin the histogram, for example
by pairing neighboring ranks together into a single bin to give
$B = L / 2$ total bins. Our experiments have shown that keeping
$N/B \approx 20$ lead to a good trade-off between the expressiveness
of the binned histogram and the necessary variance reduction.
Choosing $L + 1$ to be divisible by a large power of 2 makes this
re-binning easier; for example, instead of generating 1000 draws
in a problem with known computational limitations, one could sample $1024 - 1 = 1023$ draws
from the posterior distributions.

Regardless of the binning, however, it will be difficult to identify
sufficiently small deviations in the SBC histogram and it can be
useful to consider alternative visualizations of the rank statistics.
We consider this Section \ref{sec:small_deviations}.

\subsection{Interpreting SBC}
\label{interpretation}

What makes the SBC procedure particularly useful is that the
deviations from uniformity in the SBC histogram can indicate \emph{how}
the computed posteriors are incorrect. We follow an observation
from the forecast calibration literature
\citep{anderson1996method,hamill2001interpretation}, which suggests
that the way the rank histogram deviates from uniformity can
indicate bias or mis-calibration of the computed posterior distributions.

\begin{figure}
    \centering
    \begin{minipage}{0.45\textwidth}
        \centering
        \includegraphics[width=\textwidth]{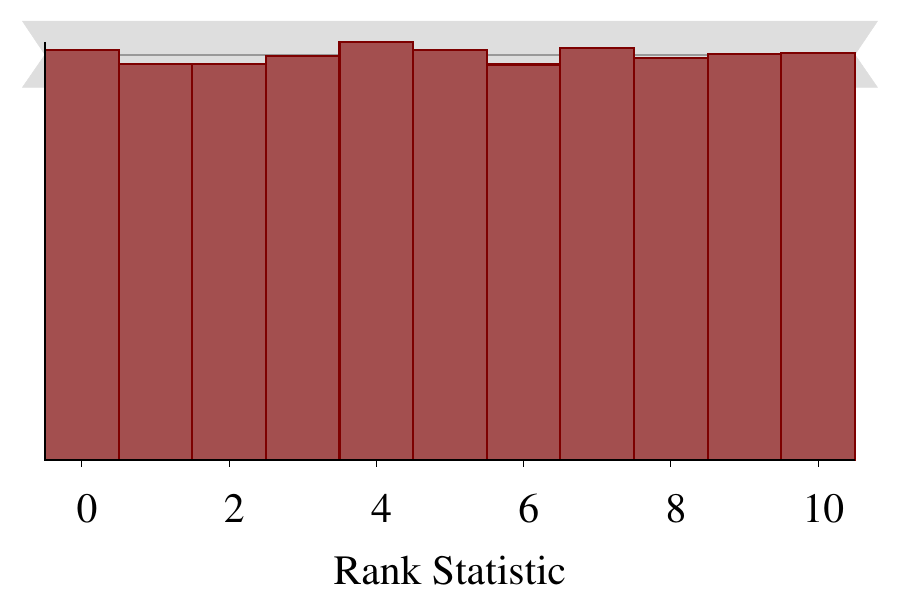}
        \caption{Uniformly distributed rank statistics are consistent
                 with the ranks being computed from independent samples
                 from the exact posterior of a correctly specified model.}
        \label{fig:rank_good}
    \end{minipage}\hfill
    \begin{minipage}{0.45\textwidth}
        \centering
        \includegraphics[width=\textwidth]{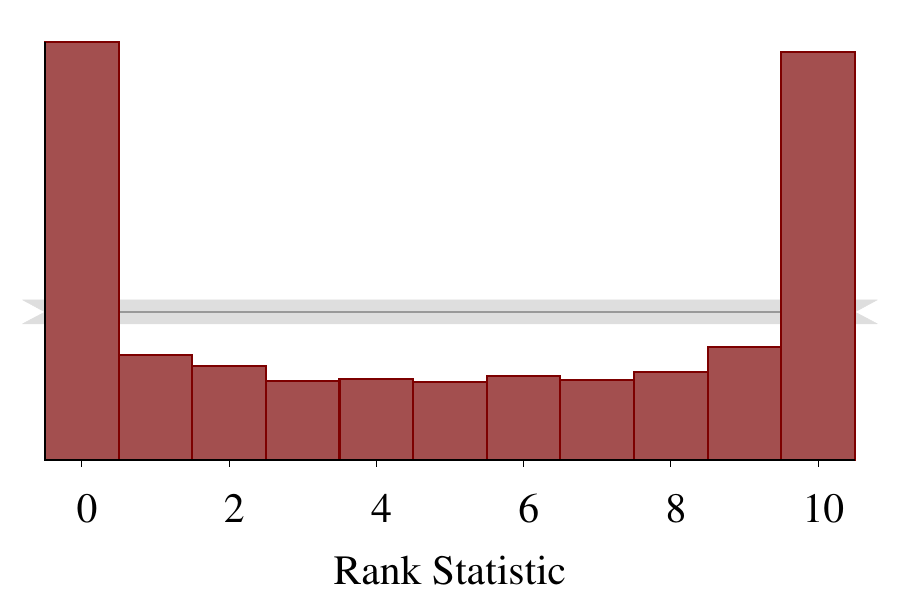}
        \caption{The spikes at the boundaries of the SBC histogram
                 indicate that posterior samples possess non-negligible
                 autocorrelation.}
        \label{fig:corr}
    \end{minipage}
\end{figure}

A histogram without any appreciable deviations is shown in Figure
\ref{fig:rank_good}. The histogram of rank statistics is consistent
with the expected uniform behavior, here shown with the 99\%
interval in light gray and the median in dark gray.

Figure \ref{fig:corr} demonstrates the deviation from uniformity
exhibited by correlated posterior samples. The correlation
between the posterior samples causes them to cluster relative
to the proceeding prior sample, biasing the ranks to extremely
small or large values. The similarity to Figure \ref{fig:lin_regr_c1}
is no coincidence. We describe how to process correlated posterior
samples generated from Markov chain Monte Carlo algorithms in
Section \ref{dealAutoCor}.

\begin{figure*}
\centering
\subfigure[]{ \includegraphics[width=2.7in]{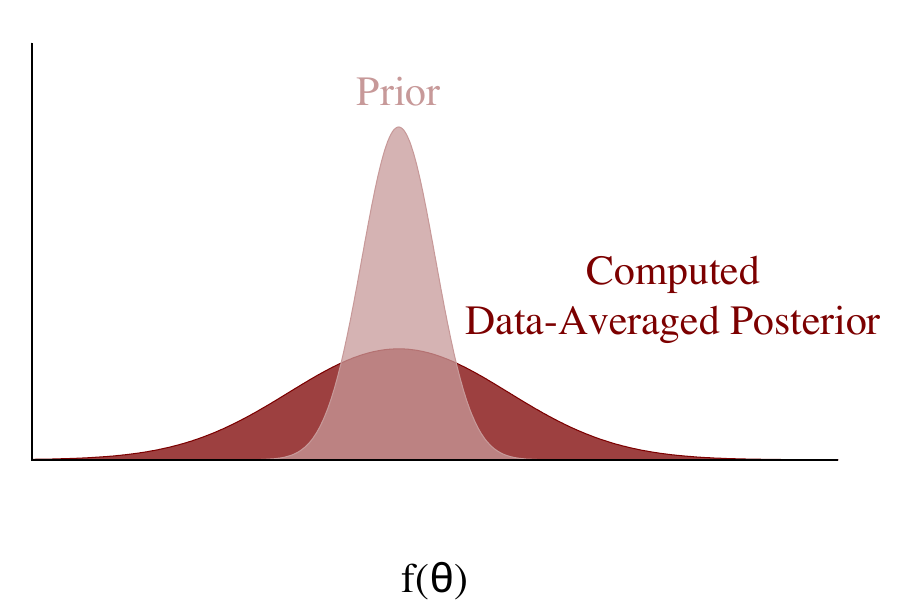} }
\subfigure[]{ \includegraphics[width=2.75in]{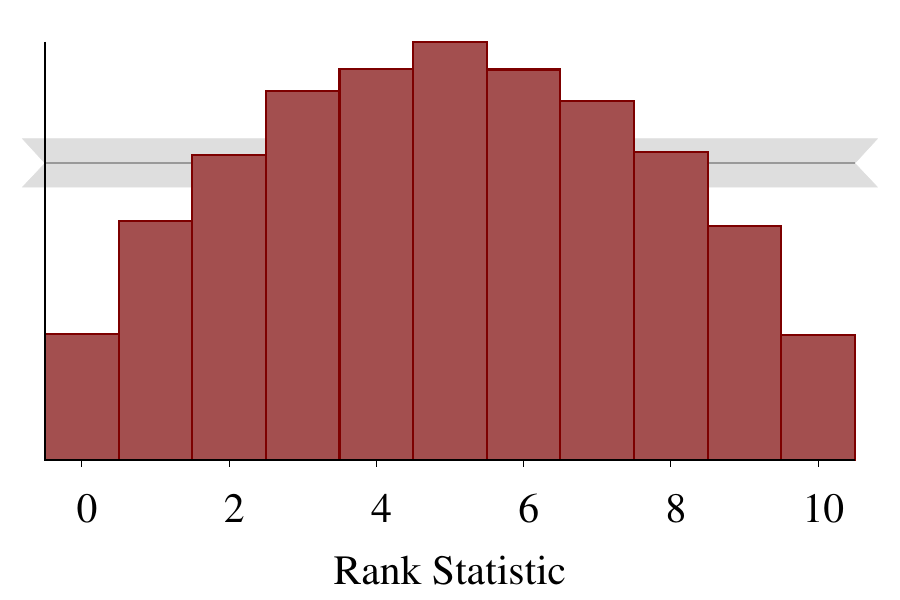} }
\caption{A symmetric, $\cap$-shaped distribution indicates that the
computed data-averaged posterior distribution (dark red) is overdispersed
relative to the prior distribution (light red). This implies that on
average the computed posterior will be wider than the true posterior.}
\label{fig:overdispersed}
\end{figure*}

Next, consider a computational algorithm that produces, on average,
posteriors that are \emph{overdispersed} relative to the true posterior.
When averaged over the Bayesian joint distribution this results in a
data-averaged posterior distribution \eqref{eqn:marginal_post} that is
overdispersed relative to the prior distribution (Figure
\ref{fig:overdispersed}a), and hence rank statistics that are biased
towards the extremes that manifests as a characteristic $\cap$-shaped
histogram (Figure \ref{fig:overdispersed}b).

\begin{figure*}
\centering
\subfigure[]{ \includegraphics[width=2.7in]{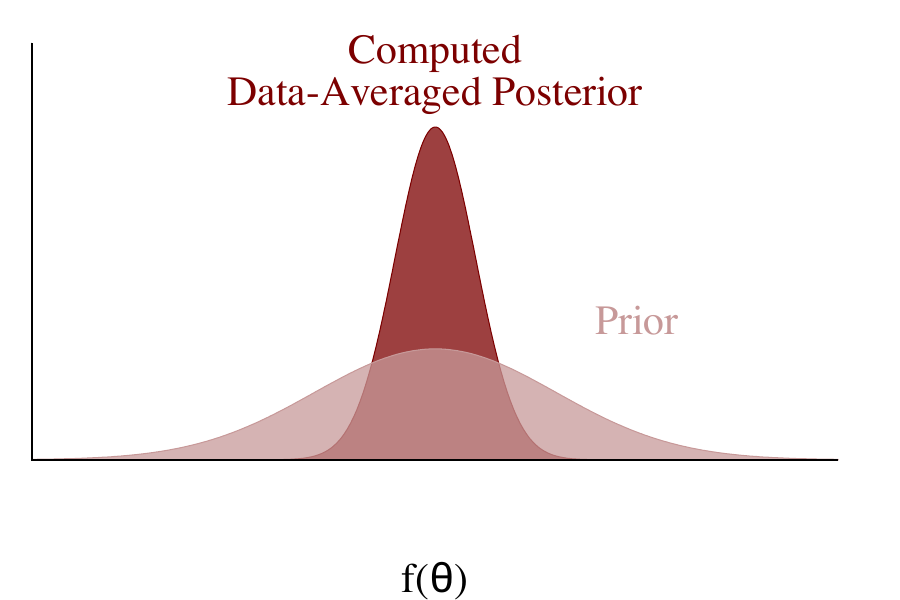} }
\subfigure[]{ \includegraphics[width=2.75in]{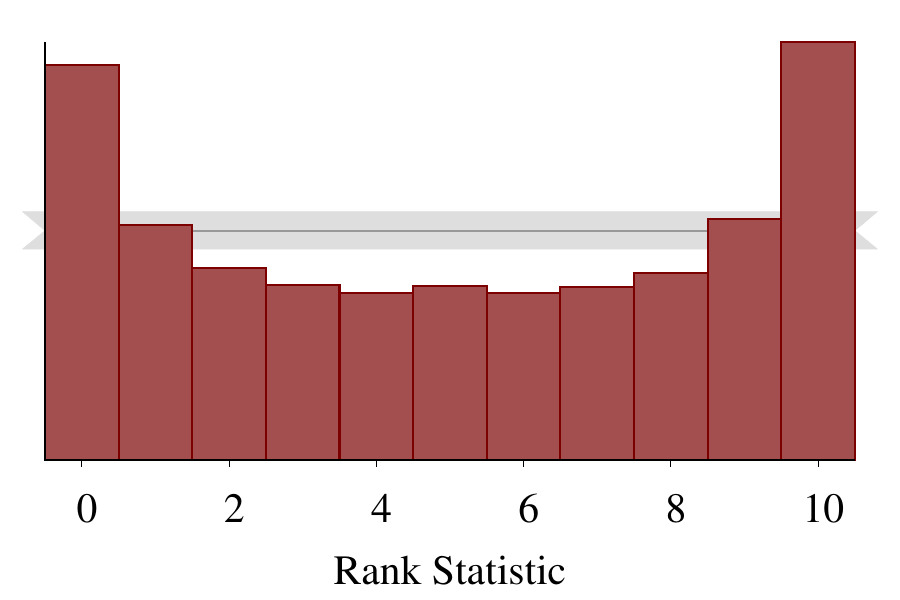} }
\caption{A symmetric $\cup$ shape indicates that the computed
data-averaged posterior distribution (dark red) is under-dispersed relative
to the prior distribution (light red). This implies that on
average the computed posterior will be narrower than the true posterior.}
\label{fig:underdispersed}
\end{figure*}

Conversely, an algorithm that computes posteriors that are,
on average, \emph{under-dispersed} relative to the true posterior
produces a histogram of rank statistics with a characteristic $\cup$
shape (Figure \ref{fig:underdispersed}).

\begin{figure*}
\centering
\subfigure[]{ \includegraphics[width=2.7in]{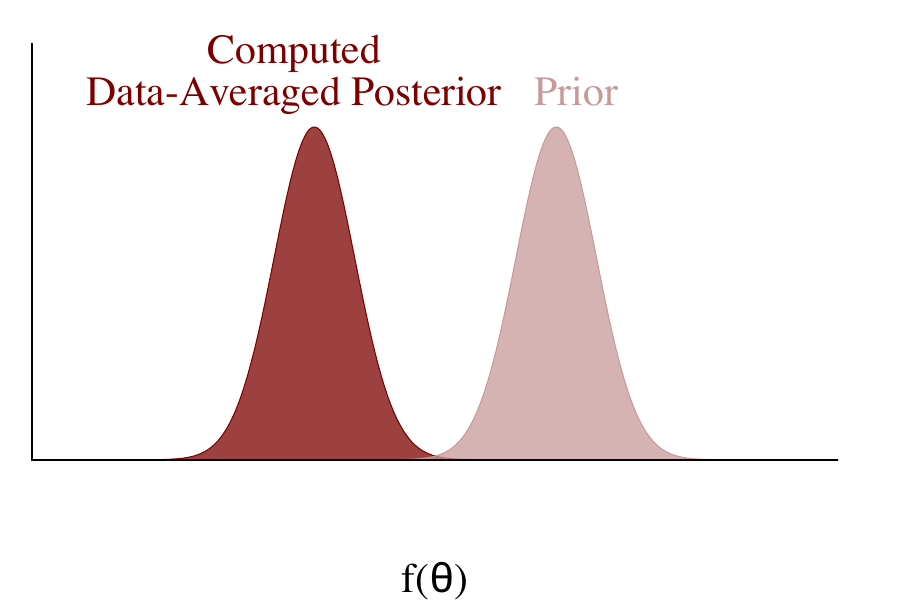} }
\subfigure[]{ \includegraphics[width=2.75in]{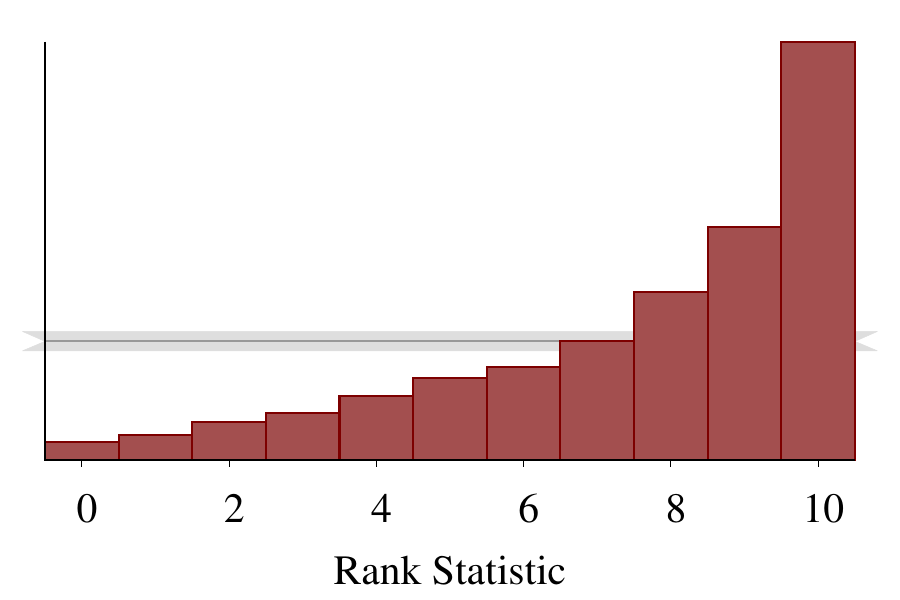} }
\caption{Asymmetry in the rank histogram indicates that the computed
data-averaged posterior distribution (dark red) will be biased in
the opposite direction relative to the prior distribution (light red).
This implies that on  average the computed posterior will be biased
in the same opposite direction.}
\label{fig:left_bias}
\end{figure*}

Finally, we might have an algorithm that produces posteriors
that are biased above or below the true posterior. This
bias results in a data-averaged posterior distribution biased
in the same direction relative to the prior distribution
(Figure \ref{fig:left_bias}a) and rank statistics that are biased
in the opposite direction (Figure \ref{fig:left_bias}b).
For example, posterior samples biased to smaller values results
in higher rank statistics, where as posterior samples biased to
larger values results in lower rank statistics.

A misbehaving analysis can in general manifest many of these
deviations at once. Because each deviation is relatively distinct
from the others, however, in practice the systematic deviations
are readily separated into the different behaviors if they are
large enough.

\subsection{Simulation-Based Calibration Plays a Vital Role
in a Robust Bayesian Workflow}

SBC is one of the few tools for evaluating the critical but
frequently unexamined choice of computational method made
in any Bayesian analysis.  We have already argued that
performance on a single simulated observation is, at best, a
blunt instrument.  Moreover, while most theoretical results only
provide asymptotic comfort, SBC adapts to the specific model
design under consideration.

Furthermore, because SBC validates accuracy through one-dimensional
random variables we can use carefully chosen random variables to
make targeted assessments of an analysis based on our inferential
needs and priorities.  As these needs and priorities change we can
run SBC again to verify the analysis anew.

The downside of using SBC in practice is that it is expensive;
instead of fitting a single observation we have to fit $N$
simulated observations before even considering the measured data.
These fits, however, are embarrassingly parallel, which makes it possible to leverage
access to computational resources through multicore personal
computers, computing clusters, and cloud computing. For
example, all of the examples in Section \ref{sec:experiments}
were run on clusters and took, at most, a few hours.

The procedure can be sped up further by reducing the number of
independent draws needed from the posterior at the cost of losing
some sensitivity.  Even a few simulations are useful to catch
gross problems in an analysis.

\section{Extending Simulation-Based Calibration}

SBC provides a straightforward procedure for validating
simulation-based algorithms applied to Bayesian analyses,
but the procedure can be limited in a few circumstances.
In this section we discuss some small modifications that
allow SBC to remain useful in some common practical
circumstances.

\subsection{Mitigating the Effect of Autocorrelation}
\label{dealAutoCor}

As we saw in Section \ref{interpretation}, SBC histograms
will deviate from uniformity if the posterior samples are
dependent, making it difficult to identify any bias in
the samples. Unfortunately this limits the utility of the
ideal SBC procedure when applied to Markov chain Monte
Carlo (MCMC) algorithms. Given the popularity of these
algorithms in practice, and the consequent need for
validation schemes, we turn now to ameliorating the
effects of autocorrelation with an appropriate thinning
scheme.

Under certain ergodicity conditions, Markov chain Monte
Carlo estimators achieve a central limit theorem,
\begin{equation*}
\frac{1}{N_{\mathrm{eff}}} \sum_{n = 1}^{N_{\mathrm{eff}}}
f(\theta_{n})
\sim \mbox{N} \! \left(
\mathbb{E}[f], \frac{ \mathbb{V}[f] }{ N_{\mathrm{eff}}[f] }
\right),
\end{equation*}
where $\mathbb{E}[f]$ is the posterior expectation of
a function $f$, $\mathbb{V}[f]$ is the variance of
$f$, and $N_{\mathrm{eff}}[f]$ is the effective sample
size for $f$,
\begin{equation*}
N_{\mathrm{eff}} [f]
=
\frac{N_\text{samp}}{1 + 2 \sum_{m = 0}^{\infty} \rho_{m}[f] },
\end{equation*}
with $\rho_{m}[f]$ the lag-$m$ autocorrelation of $f$, which we
estimate from the realized Markov chain \citep[][Ch. 11]{bda3}.  In words,
$N_\text{samp}$ correlated samples contains roughly the same
information as $N_{\mathrm{eff}}$ exact samples when estimating
the expectation of $f$.


This suggests that thinning a Markov chain by keeping only every $T$
states so that $2 \sum_{m = T}^{\infty} \rho_{m}[f] < \epsilon$
should yield a sample with negligible autocorrelation. In practice, we
have observed that when $N_{\mathrm{eff}}[f] \leq N$, thinning by
$\lceil N / N_{\mathrm{eff}}[f] \rceil$ reduces the autocorrelation
sufficiently to produce samples that are well-suited for the SBC
giving us (Algorithm \ref{algo:sbc_mcmc}). Some MCMC algorithms, such
as dynamic HMC, can produce antithetic Markov chains with
$N_{\mathrm{eff}}[f]>N$. In such cases, we suggest first thinning by
$2$, which removes the negative odd lag correlations, and then thin as
suggested above.

By carefully thinning the autocorrelated samples we should be able to
significantly reduce the $\cup$ shape demonstrated in Figure
\ref{fig:corr} and maximize the sensitivity to any remaining issues
with the model or algorithm. As the rank statistic is closely related
to cumulative distribution function $P(f(\theta) \leq f^*)$, we
suggest computing minimum $N_{\mathrm{eff}}[P]$ with $f^*$ being
empirical quantiles of $f(\theta)$ (e.g. 19 equispaced quantiles).
When running the SBC procedure over multiple quantities of interest we
suggest thinning the chain just once using the largest thinning value
determined with the above procedure over all quantities.

\begin{algorithm}
\caption{Simulation-based calibration can be applied to the
correlated posterior samples generated by a Markov chain
provided that the Markov chain can be thinned to $L$ effective
samples at each iteration.}
\label{algo:sbc_mcmc}
\begin{algorithmic}[0]
\State Initialize a histogram with bins centered around $0, \ldots, L$.
\State
\For{n in N}
  \State draw a prior sample $\tilde{\theta} \sim \pi(\theta)$
  \State draw a simulated data set $\tilde{y} \sim \pi(y \mid \tilde{\theta})$
  \State run a Markov chain for $L'$ iterations to generate the correlated
         posterior samples,
  \State \hspace{5mm} $\left\{ \theta_{1}, \ldots, \theta_{L'} \right\}
         \sim \pi(\theta \mid \tilde{y})$
  \State compute the effective sample size, $N_{\mathrm{eff}}[f]$ of
         $\left\{ \theta_{1}, \ldots, \theta_{L'} \right\}$ for the function $f$
  \If{ $N_{\mathrm{eff}}[f] < L$ }
    \State rerun the Markov for $L' \cdot L / N_{\mathrm{eff}}[f]$ iterations
  \EndIf
  \State uniformly thin the correlated sample to $L$ states and truncate any
  leftover draws at $L$
  \State compute the rank statistic
  		 $r(\{f(\theta_1), \ldots, f(\theta_L)\}, f(\tilde{\theta}))$
   		 as defined in \eqref{eqn:rank_stat}
  \State increment the histogram with
         $r(\{f(\theta_1), \ldots, f(\theta_L)\}, f(\tilde{\theta}))$
\EndFor
\State Analyze the histogram for uniformity.
\end{algorithmic}
\end{algorithm}

Although some autocorrelation will remain in a sample that has been
thinned by effective sample size, our experience has been that this
strategy is sufficient to remove the autocorrelation artifacts from
the SBC histogram. If desired, more conservative thinning strategies
such as the truncation rules of \cite{geyer1992practical} can
remove autocorrelation completely from the sample. A sample thinned
with these rules is typically much smaller than the sample achieved
by thinning based on the effective sample size, and we have not seen
any significant benefit for SBC from the increased computation time
needed for these more elaborate thinning methods to date.

Deviations that cannot be mitigated by thinning provide strong
evidence that the Markov chain Monte Carlo estimators do not
follow a central limit theorem and the Markov chains are not
adequately exploring the target parameter space.  This is
particularly useful given that establishing central limit
theorems for particular Markov chains and particular target
distributions is a notoriously challenging problem even in
relatively simple circumstances.

\subsection{Simulation-Based Calibration for Small Deviations}
\label{sec:small_deviations}

The SBC histogram provides a general and interpretable means of
identifying deviations from uniformity of the rank statistics
and hence inaccuracies in our posterior computation, at least
when the inaccuracies are large enough. For small deviations,
however, the SBC histogram may not be sensitive enough for the
deviations to be evident and other visualization strategies may
be advantageous.

One option is to bin the SBC histogram multiple times to see
if any deviation persists regardless of the binning. This
approach, however, is ungainly to implement when there are many
parameters and can be difficult to interpret. In particular,
considering multiple histograms introduces a vulnerability to
multiple testing biases.

\begin{figure*}
\centering
\subfigure[]{ \includegraphics[width=2.75in]{order_good.eps}} \\
\subfigure[]{ \includegraphics[width=2.7in]{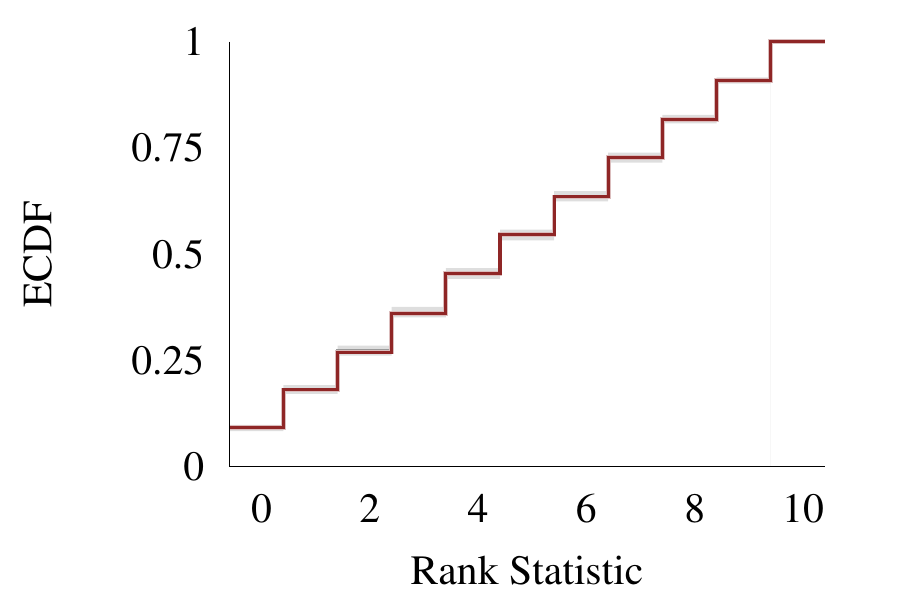} }
\subfigure[]{ \includegraphics[width=2.7in]{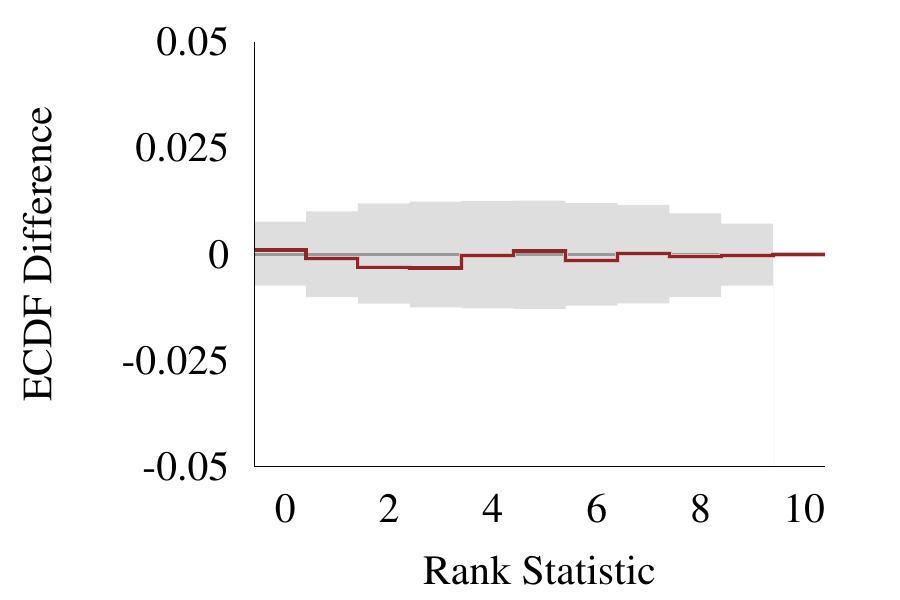} }
\caption{In order to emphasize small deviations at low and large
ranks we can pair the (a)  SBC histogram with the corresponding
(b) empirical cumulative distribution function (dark red) along
with the variation expected of the empirical cumulative distribution
function under uniformity. (c) Deviations are often easier to
identify by subtracting the expected uniform behavior from the
empirical cumulative distribution function.}
\label{fig:sbc_ecdf}
\end{figure*}

Another approach is to pair the SBC histogram with the empirical
cumulative distribution function (ECDF) which reduces
variation at small and large ranks, making it easier to identify
deviations around those values (Figure \ref{fig:sbc_ecdf}b). The
deviation of the empirical CDF away from the expected uniform
behavior is especially useful for identifying these small deviations
(Figure \ref{fig:sbc_ecdf}c).

More subtle deviations can be isolated by considering more
particular summary statistics, such as ranks quantiles or
averages. While these have the potential to identify small
biases they can also be harder to interpret and not as sensitive
to the systematic deviations that manifest in the SBC histogram.
Identifying a robust suite of diagnostic statistics is an open
area of research and at present we recommend using the SBC
histogram whenever possible.

\section{Experiments} \label{sec:experiments}

In this section we consider the application of SBC on a series
of examples that demonstrates the utility of the procedure for
identifying and correcting incorrectly implemented analyses.
For each example we implement the SBC procedure using posterior
samples $L = 100$ so that, if the algorithm is properly calibrated,
then the rank statistics will follow a $U[0,100]$ discrete uniform
distribution. The experiments in Section \ref{sec:bad_prior} through
Section \ref{advi} used $N=10,000$ replicated observations while
the experiment in Section \ref{sec:INLA} used $N=1000$ replicated
observations.

\subsection{Misspecified Prior}\label{sec:bad_prior}

Let's first consider the case where we build our posterior using a
different prior than that which we use to generate prior samples.
This is not an uncommon mistake, even when models are specified in
probabilistic programming languages.

\begin{figure*}
\includegraphics[width=2.75in]{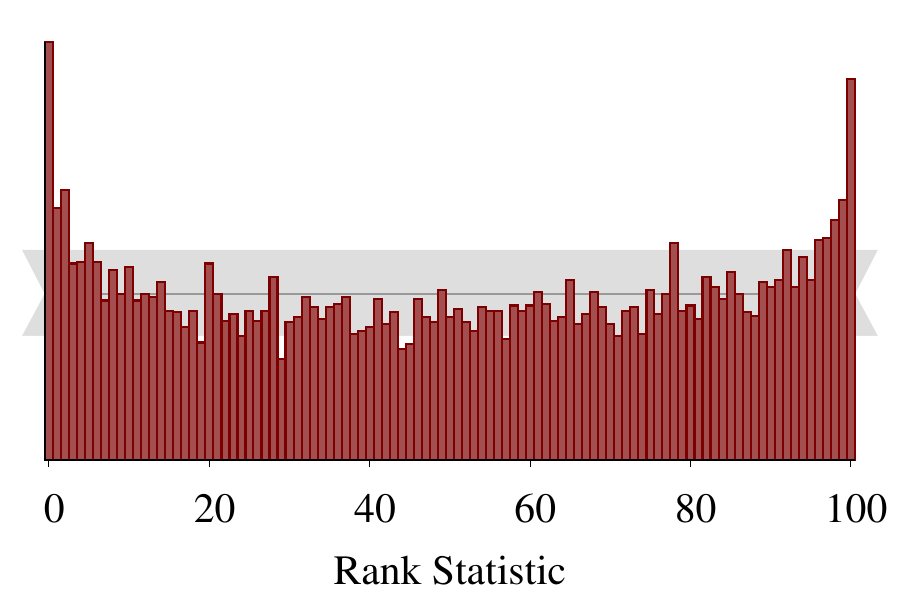}
\caption{When the data are simulated using a much wider prior than
was used to fit the model, the SBC histogram for a regression parameter
$\beta$ exhibits a characteristic $\cup$-shape.}
\label{fig:sbcwide}
\end{figure*}

Consider the linear regression model that we used before
(Listing \ref{lst:linregr_c} in the Appendix) but with the prior on
$\beta$ modified to $\mbox{N}(0, 1^{2})$. With the prior samples still
drawn according to $\mbox{N}(0, 10^2)$, we expect that the posterior
for $\beta$ will be under-dispersed relative to the prior even when
the computation is exact. This should then lead to the deviation
demonstrated in Figure \ref{fig:underdispersed} and, indeed, we see
the characteristic $\cup$ shape in the SBC histogram for $\beta$
(Figure \ref{fig:sbcwide}).

\subsection{Biased Markov chain Monte Carlo}
\label{sbc8}

Hierarchical models implemented with a centered parameterization
\citep{papaspiliopoulos2007general} are known to exhibit a
challenging geometry that can cause MCMC algorithms to return
biased posterior samples. While some algorithms, such as Hamiltonian
Monte Carlo \citep{neal2011mcmc, Betancourt2013} provide diagnostics
capable of identifying this problem, these diagnostics are not
available for general MCMC algorithms. Consequently the SBC
procedure will be particularly useful in hierarchical models if
it can identify this problem.

Here we consider a hierarchical model of the eight schools data
set \cite{Rubin1981} using a centered parameterization
(Listing \ref{lst:schoolcp} in the Appendix). In this example
the centered parameterization exhibits a classic funnel shape
that contracts into a region of strong curvature around small
values of $\tau$, making it difficult for most Markov chain
methods to adequately explore.

\begin{figure*}
\centering
\subfigure[]{ \includegraphics[width=2.7in]{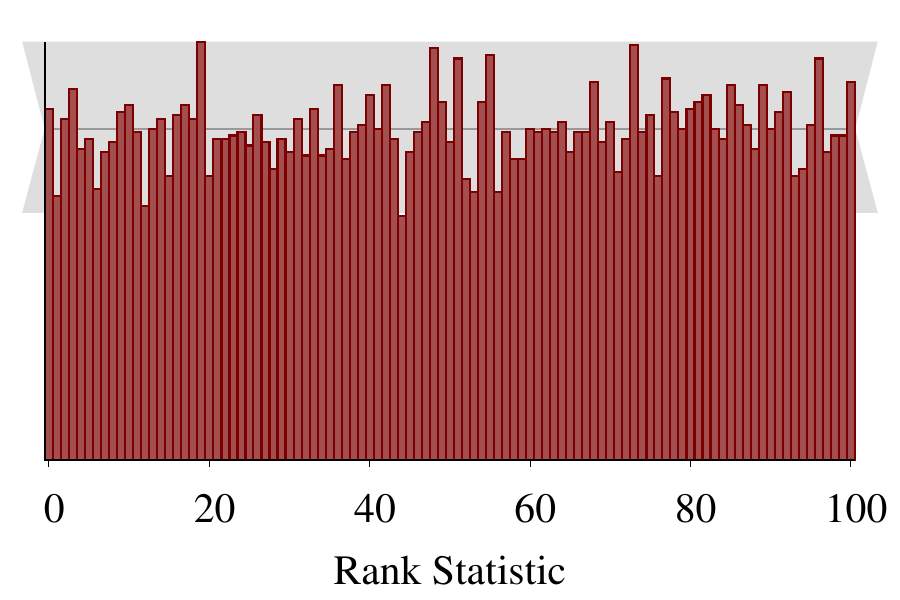} }
\subfigure[]{ \includegraphics[width=2.7in]{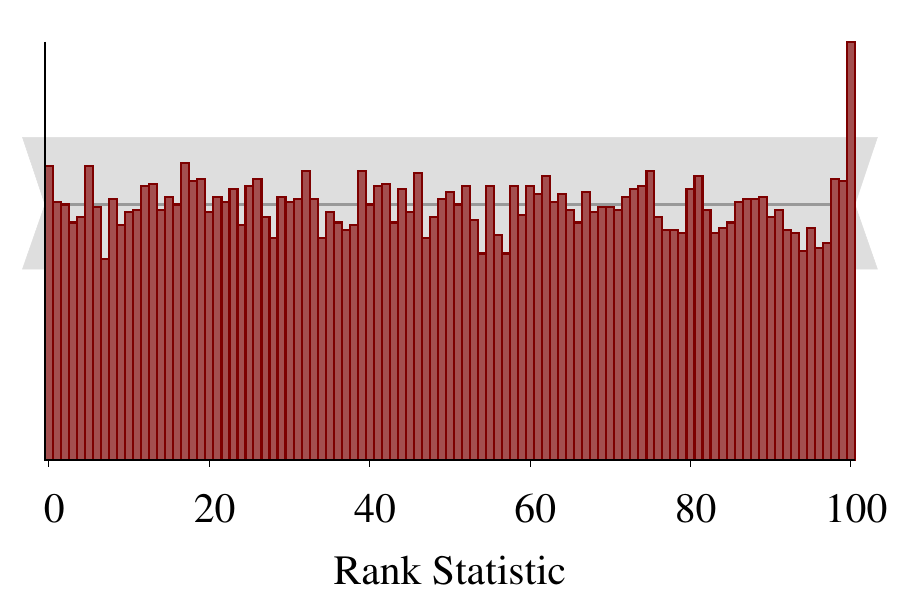} }
\caption{Once thinned (a), the SBC histogram for $\tau$ from the 8
schools non-centered parameterization in Section \ref{sbc8} show no evidence of
bias. Without thinning, the SBC histogram for $\tau$ in the same model, (b),
  exhibits characteristic signs of autocorrelation in the posterior samples.}
\label{fig:sbc8ncp}
\end{figure*}

The SBC rank histogram for $\tau$ produced from Algorithm
\ref{algo:sbc_nominal} clearly demonstrates that the posterior
samples from Stan's dynamic Hamiltonian Monte Carlo extension
of the NUTS algorithm \citep{NUTS2011,betancourt2017conceptual}
are biased below the prior samples, consistent with the known
pathology (Figure \ref{fig:sbc8ncp}b). Here we used Algorithm
\ref{algo:sbc_nominal} instead of \ref{algo:sbc_mcmc} because
the algorithm's unfaithfulness is evident over the deviation
caused by the autocorrelation. Moreover, the extra computation
required to return $L = 100$ effective samples post-thinning is
impractical here as the centered parameterization, among other
failing HMC diagnostics, has a  low effective sample size per
sample rate.

\begin{figure*}
\centering
\subfigure[]{ \includegraphics[width=2.75in]{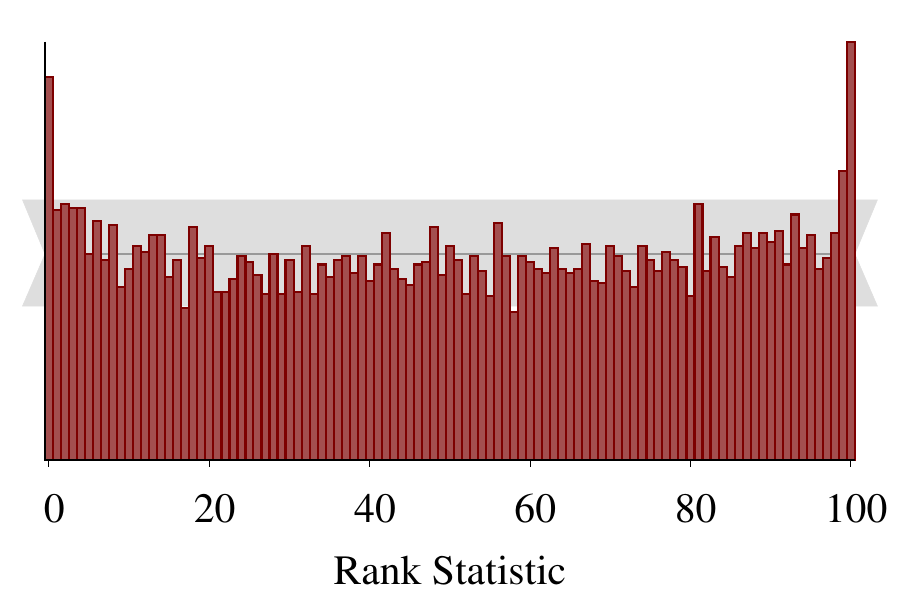}}
\subfigure[]{ \includegraphics[width=2.75in]{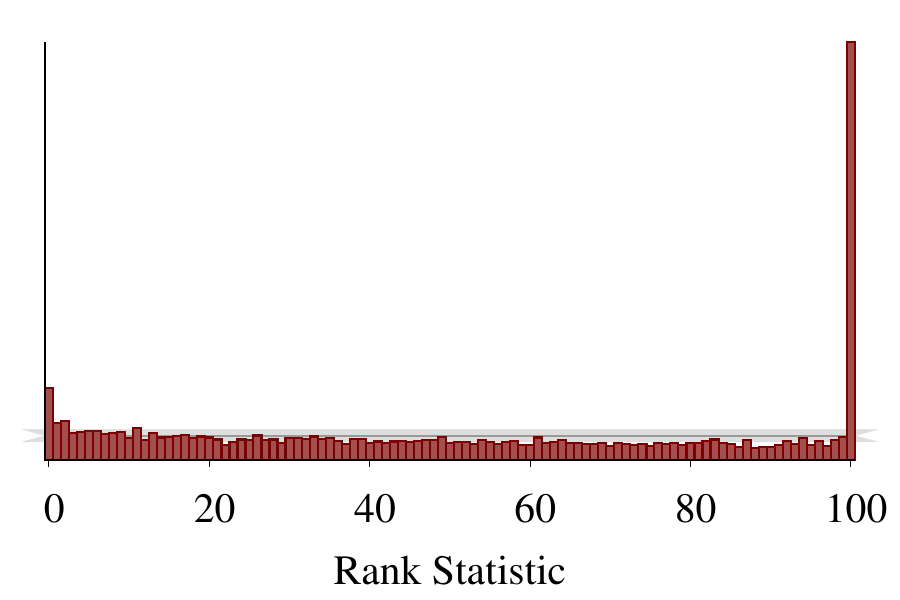} }
\caption{Even without thinning, the underlying Markov chains, the SBC
histograms for $\theta[1]$ and $\tau$ in the 8 schools centered parameterization
of Section \ref{sbc8} demonstrate that Hamiltonian Monte Carlo yields samples
that are biased towards larger values of $\tau$ than were used to generate the data.}
\label{fig:sbc8cp}
\end{figure*}

The corresponding non-centered parameterization should behave
much better. Indeed, the SBC histogram thinned using Algorithm~\ref{algo:sbc_mcmc}
(Figure \ref{fig:sbc8ncp}) shows no deviation from uniformity as
we expected given that Hamiltonian Monte Carlo is known to yield accurate
computation for this analysis. If the SBC histogram is computed without
thinning (Figure \ref{fig:sbc8ncp}), the autocorrelation manifests as a
large spikes at $L = 100$, consistent with the discussion in Section
\ref{dealAutoCor}.

\subsection{ADVI can fail for simple models}\label{advi}

We next consider automatic differentiation variational inference (ADVI)
applied to our linear regression model (Listing \ref{lst:linregr_c}
in the Appendix).  In particular, we run the implementation of ADVI in
Stan 2.17.1 that returns exact samples from a variational approximation
to the posterior.  Here we use Algorithm \ref{algo:sbc_nominal} again
because we know that ADVI does not produce autocorrelated posterior samples.

\begin{figure}
  \includegraphics[width=0.5\textwidth]{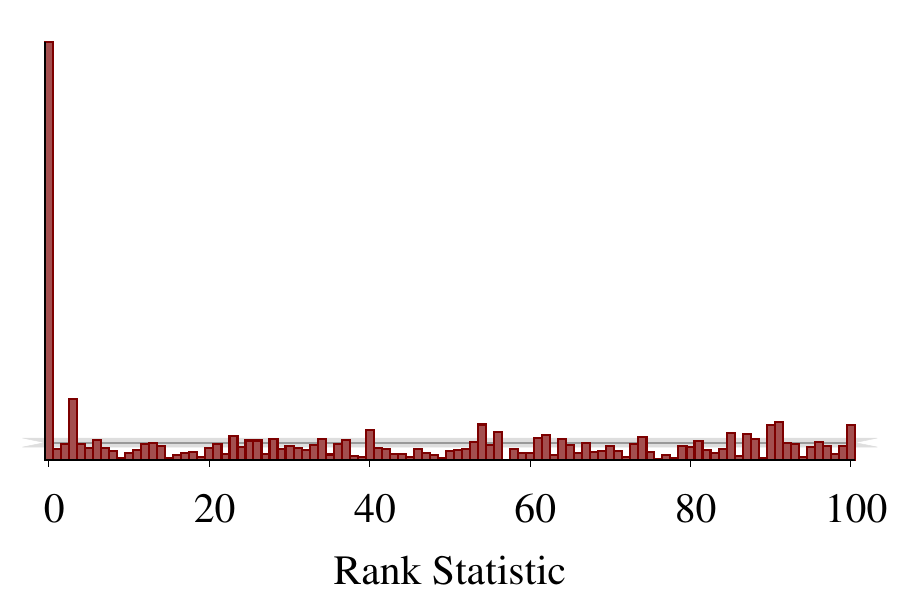}
  \caption{The SBC histogram resulting from applying ADVI on the simple
  linear regression model indicates that the algorithm is strongly biased
  towards larger values of $\beta$ in the true posterior.}
  \label{fig:lin_regr_vb}
\end{figure}

Algorithm \ref{algo:sbc_nominal} immediately identifies that the
variational approximation found by ADVI drastically underestimates
the posterior for the slope, $\beta$ (Figure \ref{fig:lin_regr_vb}).
Compare this with the results from Hamiltonian Monte Carlo (Figure
\ref{fig:sbac_lin_regr}), which yields a rank histogram consistent
with uniformity.

\subsection{INLA is slightly biased for spatial disease prevalence mapping}
\label{sec:INLA}

Finally let's consider a sophisticated spatial model for HIV prevalence
fit to data from the  2003 Demographic Health Survey in Kenya
\citep{corsi2012demographic}. We follow the experimental setup of
\citep{ wakefield2016comment} and fit the model using INLA.

The data were collected by dividing Kenya into 400 enumeration areas
(EAs) and in the $i$th EA randomly sampling $m_i$ households, with the
$j$th household containing $N_{ij}$ people.  Both $m_i$ and $N_{ij}$
are chosen to be  consistent with the Kenya DHS 2003 AIDS recode.
The number of positive responses $y_{ij}$ is modeled as
\begin{align*}
y_{ij} \sim \text{Bin}(N_{ij}, p_{ij})\\
p_{ij} = \mbox{logit}^{-1}(\beta_0 + S(x_i) + \epsilon_{ij}),
\end{align*}
where $S(\cdot)$ is a Gaussian process, $x_i$ is the centroid of the
$i$th EA, and $\epsilon_{ij}$ are iid Gaussian error terms with
standard deviation $\tau$. Following the computation reasoning of
\citet{wakefield2016comment} we approximate $S(\cdot)$ using
the stochastic partial differential equation approximation
\citep{lindgren2011explicit} to a Gaussian process with isotropic
covariance function
\begin{equation*}
c(x_1, x_2; \rho,\sigma)
=
\frac{\sqrt{8}\sigma^2}{\rho} \norm{x_1-x_2}
K_1\left(\frac{\sqrt{8}}{\rho}\norm{x_1-x_2}\right),
\end{equation*}
where $\rho$ is the distance at which the spatial correlation
between points is approximately $0.1$, $\sigma$ is the pointwise
standard deviation, and $K_1(\cdot)$ is a modified Bessel function
of the second kind.

To complete the model, we must specify priors on $\beta_0$,
$\rho$, $\sigma$, and $\tau$. We specify a $\mbox{N}(-2.5,1.5^2)$
prior on the logit baseline prevalence $\beta_0$. This prior is based on
 the national HIV prevalence across the world ranges from $0.3\%$
to $20\%$ \citep{HIV_prev}.
 We use penalized complexity priors
\citep{ simpson2017penalising,fuglstad2017constructing} on the
remaining parameters tuned to ensure
$\Pr(\rho < 0.1) = \Pr(\sigma >1) = \Pr(\tau > 1) = 0.1$.

One of the quantities of interest for this model is the average
prevalence over a subregion $A$ of Kenya,
\begin{equation*}
\frac{1}{|A|} \int_A \mbox{logit}^{-1}(\beta_0 + S(x))\,dx.
\end{equation*}
\citet{wakefield2016comment} suggested fitting this model using
the \texttt{R-INLA} package to speed up the computation. As the
quantity of interest is a non-linear transformation of a number
of parameters, we need to use the \texttt{R-INLA}'s approximate
posterior sampler, which is a relatively recent feature
\citep{seppa2017}.

\begin{figure*}
\centering
\subfigure[]{ \includegraphics[width=2.75in]{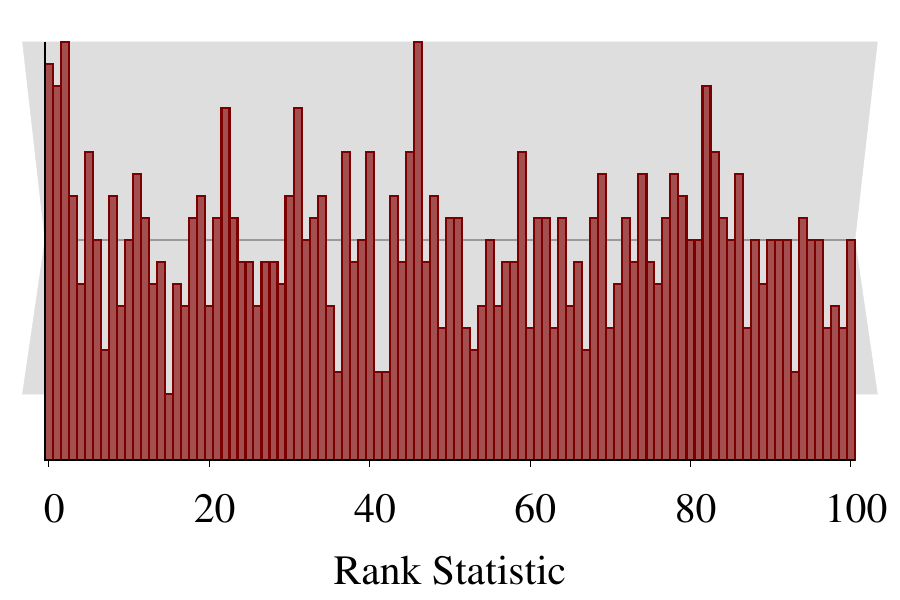}} \\
\subfigure[]{ \includegraphics[width=2.7in]{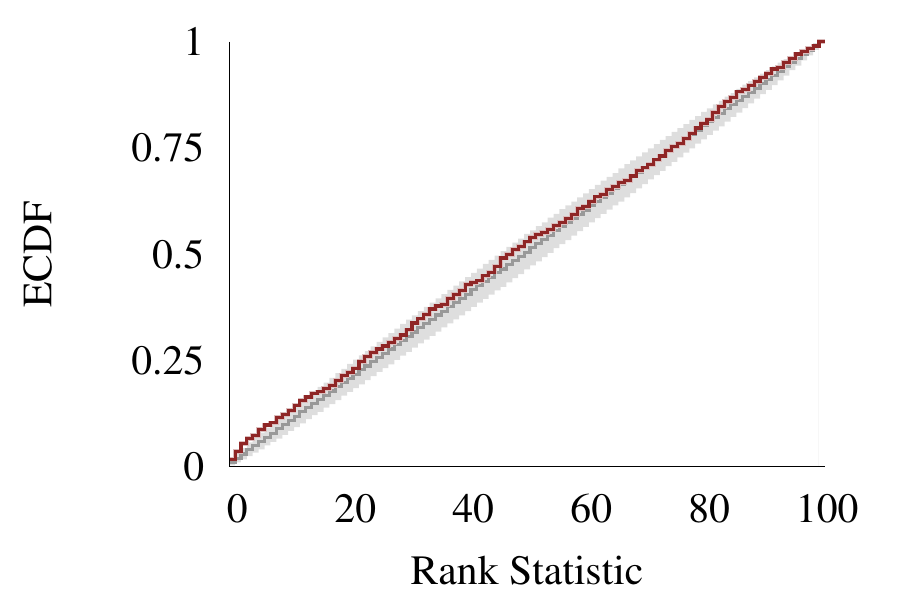} }
\subfigure[]{ \includegraphics[width=2.7in]{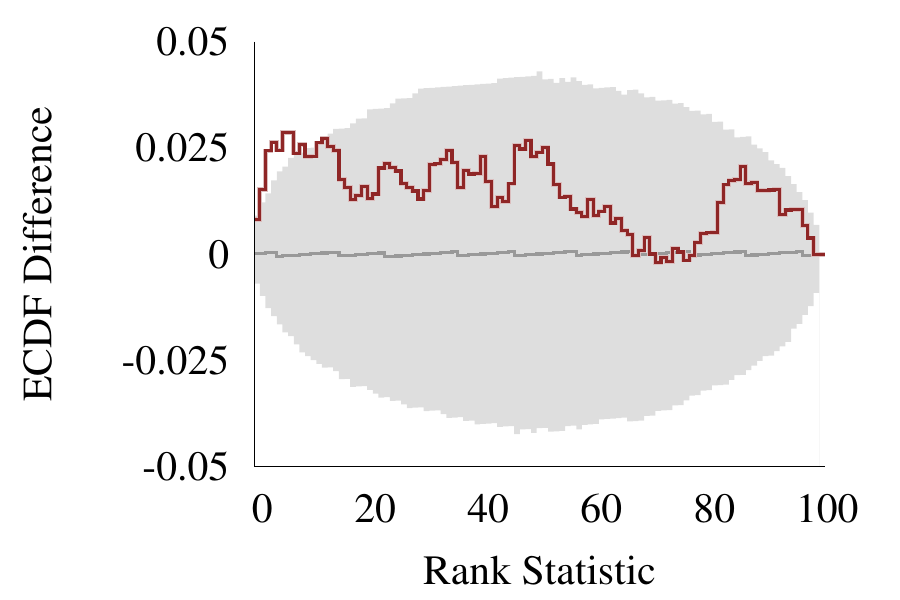} }
\caption{(a) The SBC histogram for the average prevalence of a
spatial model doesn't exhibit any obvious deviations, although
the large span of the expected variation (gray) suggests that
this test maybe too noisy to capture some potentially important discrepancies. (b) The empirical
cumulative distribution function (dark red), however, shows that
there is a small deviation at low ranks beyond the variation
expected from a uniform distribution (gray). (c) The deviation
is more evident by looking at the difference between the
empirical cumulative distribution function and the stepwise-linear
behavior expected of a discrete uniform distribution.}
\label{fig:sbc_inla}
\end{figure*}

Figure~\ref{fig:sbc_inla}a shows the SBC histogram for $N=1000$
replications to which are limited given the relatively high cost
to run INLA in this model. The histogram shows that all of the
ranks fall within the gray bars, but the large span of the bars
indicates that the visual diagnostic may be too noisy to capture some potentially important discrepancies. In our
tests, we saw that it's common for deviations from a uniform
distribution to be sufficiently severe that this histogram will
still exhibit the signs of a poorly fitting procedure. Hence for a
more fine-scale view of the fit we follow the recommendation in
Section \ref{sec:small_deviations} and consider the ECDF
(Figure~\ref{fig:sbc_inla}b, c). Here we see that low ranks are
seen slightly more often in the computed ranks than we would expect
from a uniform distribution.

It is not surprising that INLA exhibits some bias in this example.
Binomial data with low expected counts does not contain much
information, which poses some problems for the Laplace approximation.
Even though this feature is only present when the observed values of
$y_{ij}/N_{ij}$ are close to zero, the SBC procedure is a
sufficiently sensitive instrument to identify the problem. Overall,
we would view INLA as a good approximation in a country like Kenya
where the national prevalence is around $5.4\%$, while it would be
inappropriate in Australia where the prevalence is $0.1\%$ \citep{HIV_prev}.
If we repeated this type of survey in a country with only $0.1\%$
prevalence, however, then we would end up with too many zero
observations for the method to be useful.

\section{Conclusion}

In this paper, we introduce simulation-based calibration (SBC), a readily-implemented procedure
that can identify sources of poorly implemented analyses, including
biased computational algorithms or incorrect model specifications.
The visualizations produced by the procedure allow us to not only
identify that a problem exists but also learn how the problem will
affect resulting inferences.  The ability to both identify and
interpret these issues makes SBC an important step in a robust
Bayesian workflow.

Our reliance on interpreting the SBC diagnostic through visualization,
however, can be a limitation in practice, especially when dealing
with models featuring a large number of parameters. One immediate
direction for future work is to develop reliable numerical summaries
that quantify deviations from uniformity of each SBC histogram and
provide automated diagnostics that can flag certain parameters for
closer inspection.

Global summaries, such as a $\chi^2$ goodness-of-fit test of the SBC
histogram with respect to a uniform response, are natural options,
but we found they did not perform particularly well in
the above examples.  The reason for this is that the deviation from
uniformity tends to occur in only a few systematic ways, as
discussed in Section~\ref{interpretation}, whereas these tests
consider only global behavior and hence do not exploit these known
failure modes. A potential alternative is to report a number of
summaries that are designed to be sensitive to the specific types
of deviation from uniformity we might expect to see.

Another future direction is deriving the expected behavior of the
SBC histograms in the presence of autocorrelation and dropping the
thinning requirement of SBC. This could even be done empirically,
using the output of chains with known autocorrelations to calibrate
the deviations in the rank histograms. These calibrated deviations
could be used to define a sense of effective sample size for
\emph{any} algorithm capable of generating samples, not just Markov
chain Monte Carlo.

Finally, the SBC histograms are only able to assess the calibration of
one-dimensional posterior summaries. This is a limitation, especially
in situations where the quantities of interest are naturally multivariate.
An interesting extension of this methodology would be to incorporate
some of the advances in multivariate calibration of probabilistic
forecasts \citep{gneiting2008assessing,Thorarinsdottir2016}.

\subsubsection*{Acknowledgements.}
We thank Bob Carpenter, Chris Ferro, and Mitzi Morris for their helpful comments. The plot in Figure \ref{fig:sbc_inla}(c) shares the same derivation as the \texttt{inla.ks.plot} function written by Finn Lindgren and found in the R-INLA package. We thank the Academy of Finland (grant 313122), Sloan Foundation (grant G-2015-13987), U.S. National Science Foundation (grant CNS-1730414), Office of Naval Research (grants N00014-15-1-2541, N00014-16-P-2039, and N00014-19-1-2204), Defense Advanced Research Projects Agency (grant DARPA BAA-16-32),  Institute of Education Sciences (grant R305D190048), and Schmidt Futures for partial support of this research.

\bibliographystyle{imsart-nameyear}
\bibliography{references}
\appendix
\clearpage

\section{Code Listings}

We advise the reader to keep in mind that the Stan modeling language
parameterizes the normal distribution using the mean and standard
deviation whereas we have used a mean and variance parameterization
throughout this text.

\begin{lstlisting}[caption={Data generating process for linear
  regression},label={lst:gen_lin_regr_c}]
data {
  int<lower=1> N;
  real X[N];
}

generated quantities {
  real beta;
  real alpha;
  real y[N];

  beta = normal_rng(0, 10);
  alpha = normal_rng(0, 10);

  for (n in 1:N)
    y[n] = normal_rng(X[n] * beta + alpha, 1.2);
}
\end{lstlisting}
\begin{lstlisting}[caption={Inference model for linear regression}
  ,label={lst:linregr_c}]
data {
  int<lower=1> N;
  vector[N] X;
  vector[N] y;
}

parameters {
  real beta;
  real alpha;
}

model {
  beta ~ normal(0, 10);
  alpha ~ normal(0, 10);

  y ~ normal(X * beta + alpha, 1.2);
}
\end{lstlisting}
\begin{lstlisting}[caption={8 schools, centered parameterization}
  ,label={lst:schoolcp}]
data {
  int<lower=0> J;
  real y[J];
  real<lower=0> sigma[J];
}

parameters {
  real mu;
  real<lower=0> tau;
  real theta[J];
}

model {
  mu ~ normal(0, 5);
  tau ~ normal(0, 5);
  theta ~ normal(mu, tau);
  y ~ normal(theta, sigma);
}
\end{lstlisting}
\begin{lstlisting}[caption={8 schools, non-centered parameterization}
  ,label={lst:schoolncp}]
data {
  int<lower=0> J;
  real y[J];
  real<lower=0> sigma[J];
}

parameters {
  real mu;
  real<lower=0> tau;
  real theta_tilde[J];
}

transformed parameters {
  real theta[J];
  for (j in 1:J)
    theta[j] = mu + tau * theta_tilde[j];
}

model {
  mu ~ normal(0, 5);
  tau ~ normal(0, 5);
  theta_tilde ~ normal(0, 1);
  y ~ normal(theta, sigma);
}
\end{lstlisting}

\section{Proof of Theorem 1}
\label{sec:proof}

\begin{theorem} \label{theorem1}
Let $\tilde{\theta} \sim \pi(\theta)$, $\tilde{y} \sim \pi(y \mid \tilde{\theta})$,
and $\left\{ \theta_{1}, \ldots, \theta_{L} \right\}$ sampled independently from $\pi(\theta \mid \tilde{y})$
for any joint distribution $\pi(y, \theta)$. The rank statistic of any one-dimensional
random variable over $\theta$ is uniformly distributed over the integers $[0, L]$.
\end{theorem}

\begin{proof}

Consider the one-dimensional random variable $f: \Theta \rightarrow \mathbb{R}$
and let $\tilde{f} = f(\tilde{\theta})$ be the evaluation of the random variable
with respect to the prior sample with $f_{l} = f(\theta_{l})$ the evaluation of
the random variable with respect to one draw from the posterior sample.  Similarly let $\pi(f)$
and $\pi(f \mid \tilde{y})$ denote the pushforward probability density function
of the prior density function and posterior density function, respectively.

Without loss of generality we can relabel the elements of the posterior sample such
that they are ordered with respect to the random variable,
\begin{equation*}
f_{1} \le f_{2} \le \ldots \le f_{L - 1} \le f_{L}.
\end{equation*}
We can then write the probability mass function of the prior rank statistic as
\begin{align*}
\pi(r)
&=
\int \mathrm{d} f \, \mathrm{d} y \, \pi(y, f)
\frac{L!}{r!(L - r)!}
\mathbb{P} \left[ f_{l} < f \right]
\cdot
\mathbb{P} \left[ f_{l} \ge f \right]
\\
&=
\frac{L!}{r!(L - r)!}
\int \mathrm{d} f \, \mathrm{d} y \, \pi(y, f)
\mathbb{P} \left[ f_{l} < f \right]
\cdot
\mathbb{P} \left[ f_{l} \ge f \right]
\\
&=
\frac{L!}{r!(L - r)!}
\int \mathrm{d} f \, \mathrm{d} y \, \pi(y, f)
\left[ \prod_{l = 1}^{r}
\int_{-\infty}^{f} \mathrm{d} f_{l} \,
\pi(f_{l} \mid f, y)
\right]
\left[ \prod_{l = r + 1}^{L}
\int_{f}^{\infty} \mathrm{d} f_{l} \,
\pi(f_{l} \mid f, y)
\right]
\\
&=
\frac{L!}{r!(L - r)!}
\int \mathrm{d} f \, \mathrm{d} y \, \pi(y, f)
\left[
\int_{-\infty}^{f} \mathrm{d} f_{l} \,
\pi(f_{l} \mid f, y)
\right]^{r}
\left[
\int_{f}^{\infty} \mathrm{d} f_{l} \,
\pi(f_{l} \mid f, y)
\right]^{L - r}
\\
&=
\frac{L!}{r!(L - r)!}
\int \mathrm{d} f \, \mathrm{d} y \, \pi(y, f)
\left[
\int_{-\infty}^{f} \mathrm{d} f_{l} \,
\pi(f_{l} \mid f, y)
\right]^{r}
\left[ 1 -
\int_{-\infty}^{f} \mathrm{d} f_{l} \,
\pi(f_{l} \mid f, y)
\right]^{L - r}
\end{align*}

Once we condition on an observation the distribution of the posterior samples is 
independent of the conditioning model configuration,
\begin{equation*}
\pi(f_{l} \mid f, y) = \pi(f_{l} \mid y).
\end{equation*}
Consequently
\begin{align*}
\pi(r)
&=
\frac{L!}{r!(L - r)!}
\int \mathrm{d} f \, \mathrm{d} y \, \pi(y, f)
\left[
\int_{-\infty}^{f} \mathrm{d} f_{l} \,
\pi(f_{l} \mid y)
\right]^{r}
\left[ 1 -
\int_{-\infty}^{f} \mathrm{d} f_{l} \,
\pi(f_{l} \mid y)
\right]^{L - r}
\\
&=
\frac{L!}{r!(L - r)!}
\int \mathrm{d} f \, \mathrm{d} y \, \pi(f \mid y ) \, \pi(y)
\left[
\int_{-\infty}^{f} \mathrm{d} f_{l} \,
\pi(f_{l} \mid y)
\right]^{r}
\left[ 1 -
\int_{-\infty}^{f} \mathrm{d} f_{l} \,
\pi(f_{l} \mid y)
\right]^{L - r}
\\
&=
\frac{L!}{r!(L - r)!}
\int \mathrm{d} y \, \pi(y)
\int \mathrm{d} f \, \pi(f \mid y )
\left[
\int_{-\infty}^{f} \mathrm{d} f_{l} \,
\pi(f_{l} \mid y)
\right]^{r}
\left[ 1 -
\int_{-\infty}^{f} \mathrm{d} f_{l} \,
\pi(f_{l} \mid y)
\right]^{L - r}
\end{align*}

Now because the model used to simulate data and construct posterior distributions 
is the same we have
\begin{equation*}
\pi(f_{l} \mid y) = \pi(f\mid y ).
\end{equation*}
This allows us to consider the change of variables
\begin{equation*}
u(y) = \int_{-\infty}^{f} \mathrm{d} f' \,
\pi(f' \mid y)
\end{equation*}
which gives
\begin{align*}
\pi(r)
&=
\frac{L!}{r!(L - r)!}
\int \mathrm{d} y \, \pi(y)
\int \mathrm{d} u
\left[ u \right]^{r}
\left[ 1 - u \right]^{L - r}
\\
&=
\frac{L!}{r!(L - r)!}
\int \mathrm{d} y \, \pi(y)
\frac{r!(L - r)!}{(L + 1)!}
\\
&=
\frac{L!}{r!(L - r)!}
\frac{r!(L - r)!}{(L + 1)!}
\int \mathrm{d} y \, \pi(y)
\\
&=
\frac{1}{L + 1}
\int \mathrm{d} y \, \pi(y)
\\
&=
\frac{1}{L + 1},
\end{align*}
consistent with a uniform distribution over the $L + 1$ possible ranks, as desired.

\end{proof}

\end{document}